Spatiotemporal Network Evolution of Anthropogenic Night Light 1992-2015

Christopher Small

*Lamont Doherty Earth Observatory*
*Columbia University*
*Palisades, NY 10964 USA*

csmall@columbia.edu

## Abstract

Stable night light provides one of the most conspicuous indicators of human settlement and landscape modification. Satellite-derived observations of night light provide a continuous global record of lighted development extending from 1992 to the present. Global night light products, like those provided by DMSP-OLS and VIIRS dnb, provide a unique perspective on the spatial structure, and temporal evolution, of the human habitat. While not all human settlements can be detected with night light, the continuum of built environments where a rapidly growing majority of the world's population lives and works generally can. The process of segmenting a continuous luminance field into discrete spatially contiguous subsets of pixels (segments) produces a spatial network in which each contiguous segment represents a distinct network component (subset of linked nodes). Representing the continuum of lighted development as spatial networks of varying spatial connectivity allows the generative conditions for power law network structure to be applied in a spatial context to provide a general explanation for similar power law scaling observed in settlement patterns and other land cover types. This study introduces a novel methodology to combine two complementary sources of satellite-derived night light observations and use the combined product to quantify the evolution of the global spatial network structure of lighted development from 1992 to 2015. Area-perimeter distributions of network components show multifractal scaling for all years and thresholds with larger components becoming increasingly fractal. Rank-size distributions of network components are strongly linear and well described by power laws with exponents within ±0.08 of −1 for all 27 subsets of geography, year and degree of connectivity - indicating robust scaling properties. Area distributions of luminance within network components show an abrupt transition from strongly skewed to low luminance for smaller components to nearly uniform luminance distributions for larger components, suggesting a fundamental change in network structure despite a continuum of size and shape. Together, these results suggest that city size scaling, observed inconsistently for administratively defined city populations, is more consistent for physically defined settlement network area and can be explained more simply as a spatial network growth process.

# Introduction

Stable night light provides one of the most conspicuous indications of human settlement and landscape modification. Even at the global scale, where only the largest and brightest lights are resolved, it is apparent that the spatial distribution of night light spans a wide range of environments, yet remains strongly clustered on a range of spatial scales (Figure 1). The spatial distribution and interconnectedness of different sizes of settlement provide fundamental observations on the form and function of human modified landscapes. Satellite observations of anthropogenic night light provide a global measure of the spatiotemporal evolution of lighted development over the past 28 years. Characterization and identification of systematic patterns of past spatiotemporal evolution may allow for anticipation of future evolution.

Anthropogenic night light arises from a variety of sources. Sub-meter night light imagery of Berlin [*Kuechly et al.*, 2012; *Kyba et al.*, 2015], Birmingham [*Hale et al.*, 2013], Brisbane [*Levin et al.*, 2014], Hangzhou [*Zheng et al.*, 2018] and Las Vegas [*Kruse and Elvidge*, 2011] consistently show diffuse scattering from lighted roads and streets to be the primary source of nocturnal luminance in urban settings. Lighted outdoor spaces in commercial, industrial and residential settings contribute, but comparisons of astronaut photos of night lights and daytime satellite imagery of urban environments show transportation networks as the primary source of lighted area [*Small*, 2019]. While the vast majority of anthropogenic night light is associated with outdoor lighting of settlements, a significant contribution is associated with hydrocarbon production – both from flaring of gas and from lighting of production facilities. Light emitted by fires, both natural and anthropogenic is also observed in a wide range of environments, but fires are temporally intermittent and therefore easily distinguished from more stable sources associated with lighted infrastructure. Stable night light is considered uniquely anthropogenic because other sources of luminance (e.g., fire, bioluminescence, volcanic eruptions, lightning, aurora) are intermittent on time scales of seconds to days.

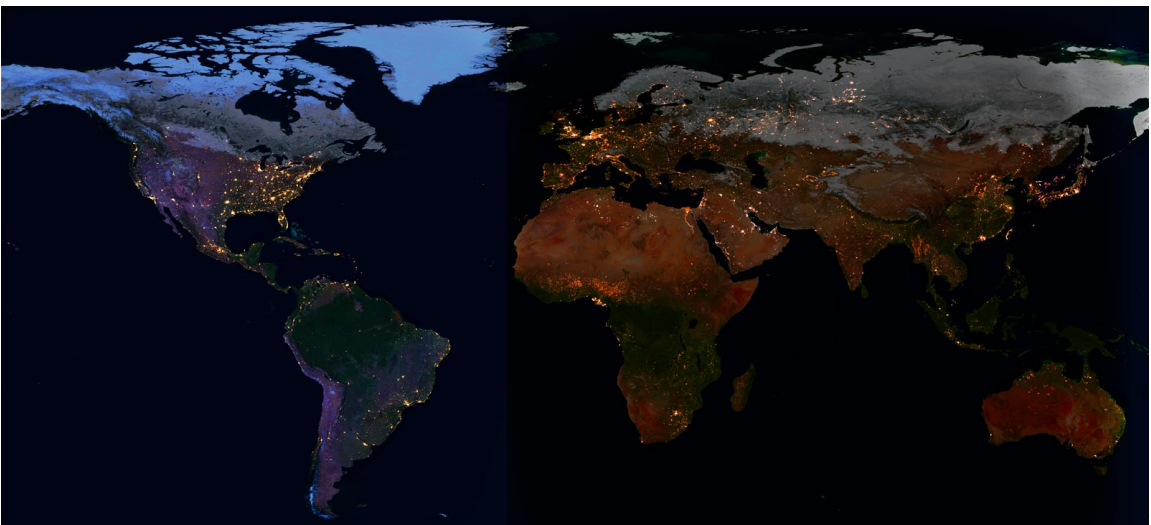

Figure 1 Anthropogenic night light and land cover. VIIRS dnb night light is superimposed on MODIS land surface reflectance. As only the largest and brightest sources are visible at global scale, the true extent of human habitation is considerably greater than indicated here.

Satellite-derived observations of night light provide a continuous global record of lighted development extending from 1992 to the present. The Defense Meteorological Satellite Program (DMSP) series of satellites collected digital nighttime imagery between 1992 and 2013 – during which time annual mean brightness composites of stable night light were produced by the Earth Observation Group at the National Geophysical Data Center (described in more detail below). Since 2012, the Visible Infrared Imaging Radiometer Suite (VIIRS) sensor has been collecting nighttime imagery. In comparison to the DMSP Operational Line Scan (OLS) sensor, the VIIRS sensor provides greater dynamic range (6 vs 14 bit), higher and more consistent spatial resolution (0.75 vs >2.7 km), on-board calibration and improved positional accuracy. Because of these improvements, VIIRS resolves two of the principal challenges of working with DMSP imagery. The greater dynamic range and higher spatial resolution enables VIIRS to image smaller, dimmer lights than OLS while also resolving luminance variations within bright urban cores, for which the OLS sensors consistently saturate. VIIRS' higher spatial resolution, improved positional accuracy and improved sensor design also considerably reduce the "overglow" effect, which causes DMSP-OLS to image night lights as being significantly larger in area than the actual source – effectively creating a halo that tends to blur the actual shape of the source. While several studies have attempted to minimize the overglow effect using various image sharpening approaches, the results are generally nonunique and do not approach the level of detail provided by VIIRS. So, while VIIRS provides superior quality imagery, it has only been doing so since 2012, while DMSP-OLS provides a multi-decadal time series of annual composites, but with considerably less spatial detail and dynamic range than VIIRS.

Global night light products, like those provided by DMSP-OLS and VIIRS, provide a unique perspective on the spatial structure, and temporal evolution, of the human habitat. While not all human settlements can be detected with night light, the continuum of built environments where a rapidly growing majority of the world's population lives and works generally can. In addition to accelerating rural-urban migration, rural electrification continues to extend anthropogenic night light to smaller and more remote settlements. In a spatial analysis of a 2008 DMSP-OLS composite [*Small et al.*, 2011] showed that the global rank-size distribution of night lights is well-described as a power law with a scaling exponent consistently near –1. This result is significant because it suggests that Zipf's Law of city size distributions [*Auerbach*, 1913; *Lotka*, 1941; *Zipf*, 1949] based on population has a spatial analog based on area of lighted development. However, unlike Zipf's Law based on administratively defined cities, the largest contiguous lighted areas observed by OLS are vastly larger than individual cities, rather representing interconnected spatial networks of lighted development [*Small et al.*, 2011]. The implication is that the rank-size scaling initially observed for individual settlement populations extends naturally to include interconnected spatial networks of settlements with consistent scaling properties extending from regional to global scales. Treating settlements as spatial networks allows the generative conditions for power law network structure [*Barabási and Albert*, 1999] to be applied in a spatial context to provide a general explanation for similar power law scaling observed in other land cover types [*Small and Sousa*, 2015].

The objectives of this study are to introduce a novel methodology to combine DMSP and VIIRS night light observations and use the combined product to quantify the evolution of the spatial network structure of lighted development from 1992 to 2015.  The methodology applies multiple luminance thresholds to a 2015 VIIRS night light composite to produce a set of spatial masks which are applied to DMSP-OLS night light composites from 1992, 2002 and 2013.  Using multiple luminance thresholds produces spatial networks of varying size and connectivity corresponding to different densities of development.  The masks also drastically reduce the overglow effect and project the decadal changes in spatial extent and brightness measured by DMSP-OLS onto the more detailed lighted extents of the 2015 VIIRS.  The result is a set of three lighted extents of varying connectivity for each of the three years for each of three geographic quatrospheres.  Each of these masked extents is segmented into a set of discrete spatially contiguous clusters.  The area and perimeter distributions of these clusters characterize the evolving morphology and spatial network structure of anthropogenic night light between 1992 and 2013, representing the global spatiotemporal evolution of the human habitat coincident with the largest mass migration in history.

## Data

 The Defense Meteorological Satellite Program (DMSP) series of satellites have been collecting nighttime imagery since the 1970s. Before 1992, the data were collected on film. Since 1992, a series of five DMSP missions have provided continuous digital imaging of night light. The Operational Line Scan (OLS) sensors on the DMSP satellites provide no on-board calibration, so cross calibration coefficients, derived from simultaneous imaging of stable night lights by successive generations of sensors during overlapping years, have been derived by [*Elvidge et al.*, 2009]. Intercalibration among OLS sensors on different satellites provides a 21 year time series of annual global composites of stable night light between 1992 and 2013. The OLS sensors image emitted visible and thermal infrared radiance in a 3000 km swath twice per night. Raw measurements having a ground sample distance of 0.56 km are spatially averaged to provide an effective spatial resolution 2.7 km.   Multiple cloud-free acquisitions are averaged at a grid resolution of 30 arc seconds (1 km at the Equator) to give mean annual luminance. Although the sensor was designed to image moonlit clouds, images are used from early evening (7–10 pm) acquisitions on non-moonlit nights for most research applications.  More detailed descriptions of the DMSP composite products are given by [*Elvidge et al.*, 1997; *Elvidge et al.*, 2001].

The Visible Infrared Imaging Radiometer Suite (VIIRS) sensor was launched on board the NASA-NOAA Suomi satellite in 2011. The day/night band (DNB) of the sensor collects low light imagery in a 3000 km swath at a fixed resolution of 500 m with an equator overpass time of 1 AM local time. Like DMSP night lights, individual VIIRS acquisitions are composited to exclude clouds and intermittent sources like fires. In comparison to DMSP, VIIRS provides higher dynamic range, on-board calibration, and multiple optical bands that can be used to distinguish different light sources. More detailed descriptions of the data, products, and applications of VIIRS imagery are given by [*Elvidge et al.*, 2013] and [*Miller et al.*, 2013]. The DMSP and VIIRS stable night light composites used in this study were produced by the Earth Observation Group at the Colorado School of Mines (https://payneinstitute.mines.edu/eog/)

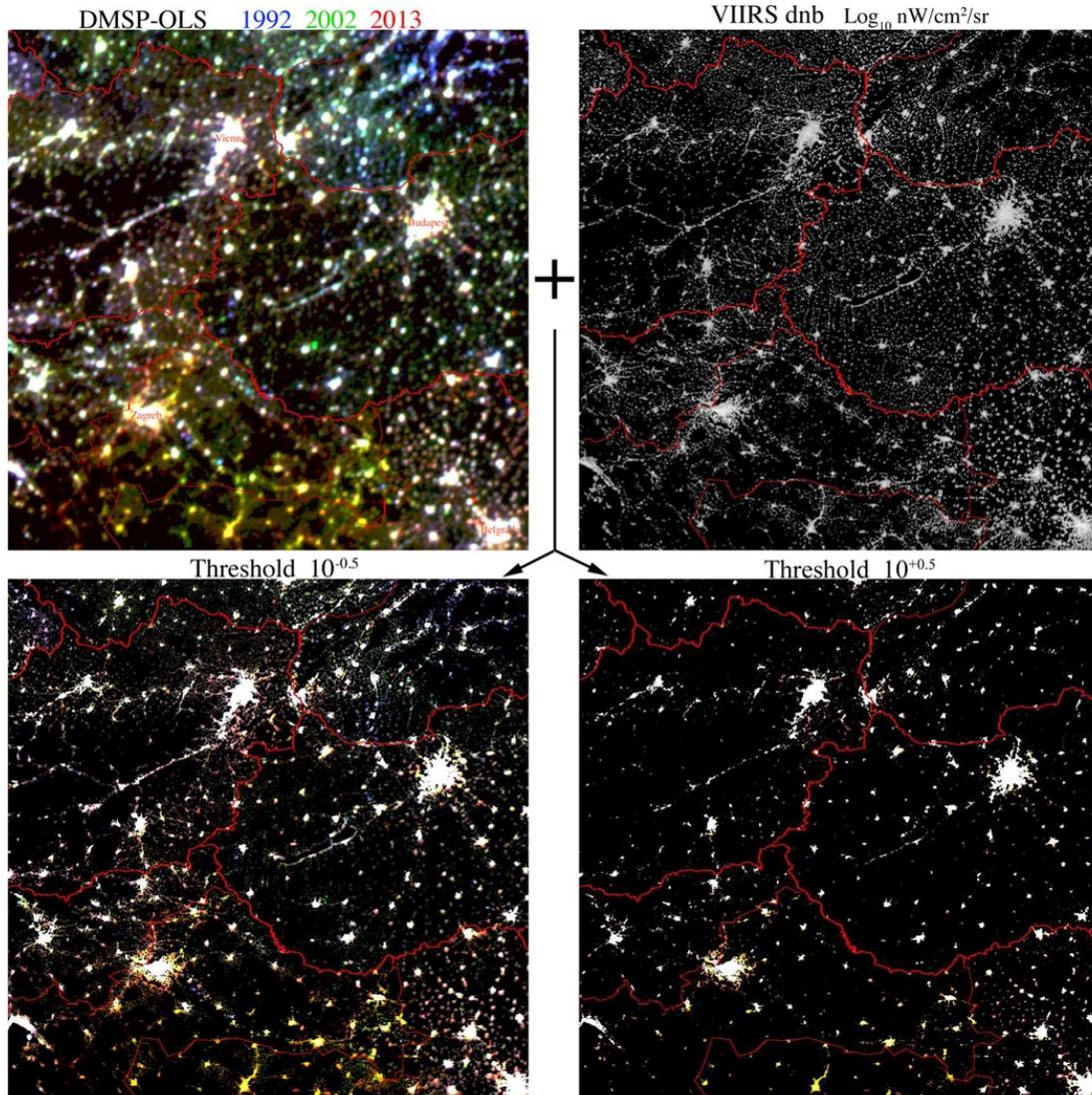

Figure 2 Threshold masking of DMSP tri-temporal composite with VIIRS dnb produces 3 tri-temporal maps of decadal change from DMSP masked to 3 lighted extents from 2015 VIIRS. Only bounding threshold extents shown. Color implies change.

## Analysis

Two combined DMSP+VIIRS products are used in this analysis. A fused DMSP+VIIRS product is used to illustrate decadal changes in DMSP night light brightness between 1992, 2002 and 2013 projected onto the more spatially detailed lighted extents from VIIRS in 2015. The fusion is effectively a pan sharpening process in which a HSV color transformation is applied to a tri-temporal RGB color composite of 2013, 2002 and 1992 DMSP luminance images. The HSV-transformed DMSP composite is coregistered and resampled to the 2015 VIIRS composite and the normalized VIIIRS luminance is used to replace the V channel of transformed DMSP composite. When the DMSP+VIIRS HSV is inverse transformed back to RGB color space the result combines the decadal change information from DMSP with the more detailed VIIRS night light. The DMSP+VIIRS fusion effectively eliminates the overglow effect by projecting the DMSP change information onto the higher spatial resolution of the 2015 VIIRS lighted extent. Because the DMSP luminance component is replaced with the more detailed VIIRS luminance, the spatial detail lost to saturation in the DMSP is effectively restored with the higher dynamic range VIIRS. The DMSP tri-temporal composite is stretched prior to HSV transformation to emphasize low luminance changes at the expense of increased saturation in the less variable urban cores. Therefore, the resulting luminance in each year is a combination of the stretched DMSP luminance and the 2015 VIIRS luminance. This fused product is used only to illustrate the 1992 to 2013 changes projected onto the more detailed 2015 VIIRS.

A separate set of masked DMSP+VIIRS products is used to quantify decadal changes in DMSP night light within the lighted extents of the 2015 VIIRS annual composite. For this product, a set of three low luminance thresholds are applied to the 2015 VIIRS composite to produce three masks which are applied to each of the individual year DMSP composites. The VIIRS masks eliminate the DMSP overglow, but do not restore the information lost to DMSP saturation. Some residual overglow is still present within the VIIRS masked extents, but the vast majority of overglow is removed by the much higher spatial resolution VIIRS extents. This process is illustrated in Figure 2. It is important to emphasize that this masked product does not change the information content of the lower resolution DMSP composites. It merely removes most of the overglow extent and projects the lower resolution luminance changes onto the more detailed 2015 lighted extents produced by the thesholding process.

Previous analyses of DMSP night light have often applied a low luminance threshold in an attempt to reduce the overglow effect. While masking pixels below the low luminance threshold can reduce overglow extent, it also attenuates smaller, dimmer light sources unrelated to bright source overglow. Aside from the loss of low luminance information, the principal difficulty with thresholding is the determination of an appropriate threshold. In a comparative analysis of DMSP composites with Landsat-derived built extents from a diverse set of 16 urban systems, [*Small et al.*, 2005] found no single luminance threshold that produced lighted extents consistent with built extents for a majority of the cities analyzed. The lack of a single universal threshold is a combined result of varying brightness gradients and varying urban morphologies. In a more recent comparative analysis of the VIIRS 2015 annual composite with multitemporal Sentinel 2 imagery for 12 urban systems, [*Small*, 2019] found VIIRS overglow effect to be considerably less than DMSP, but still sufficiently variable as to preclude the use of a single low luminance threshold for different settlement types.

Despite some overglow effect on steep luminance gradients, VIIRS' higher spatial resolution and greater dynamic range does provide considerably more information on the source of night light than DMSP. The comparison between the 2015 VIIRS product and the 10 m resolution Sentinel 2 imagery reveals a systematic relationship between the density of built surface (roads & buildings) and night light brightness. Because different luminance levels in the VIIRS composite generally correspond to different building and road density, different luminance thresholds encompass different types of development. Higher thresholds limit extent to the brightest urban cores with the highest density of outdoor lighting, while lower thresholds also encompass lower density periurban areas where lighting is more intermittent with dimmer, more diffuse sources. For this reason, different luminance thresholds can be considered representative of different ranges of development density, with lower thresholds encompassing a wider range of built environments. Given the differing implications of different luminance thresholds, it makes sense to treat the threshold itself as a variable in the analysis. In this study, three low luminance thresholds are used to produce three VIIRS masks to apply to the 1992, 2002 and 2013 DMSP composites. Figure 3 shows luminance histograms for the eastern and western hemisphere subsets of the 2015 VIIRS composite. The lower threshold ($10^{-0.5}$ nW/cm$^2$/sr) cuts only the lower tail of dimmest pixels, while the upper threshold ($10^{+0.5}$ nW/cm$^2$/sr) cuts the primary mode of the distribution and retains only the shoulder and upper tail of brighter pixels. The intermediate threshold ($10^{0}$ nW/cm$^2$/sr) retains approximately half of the lighted pixels.

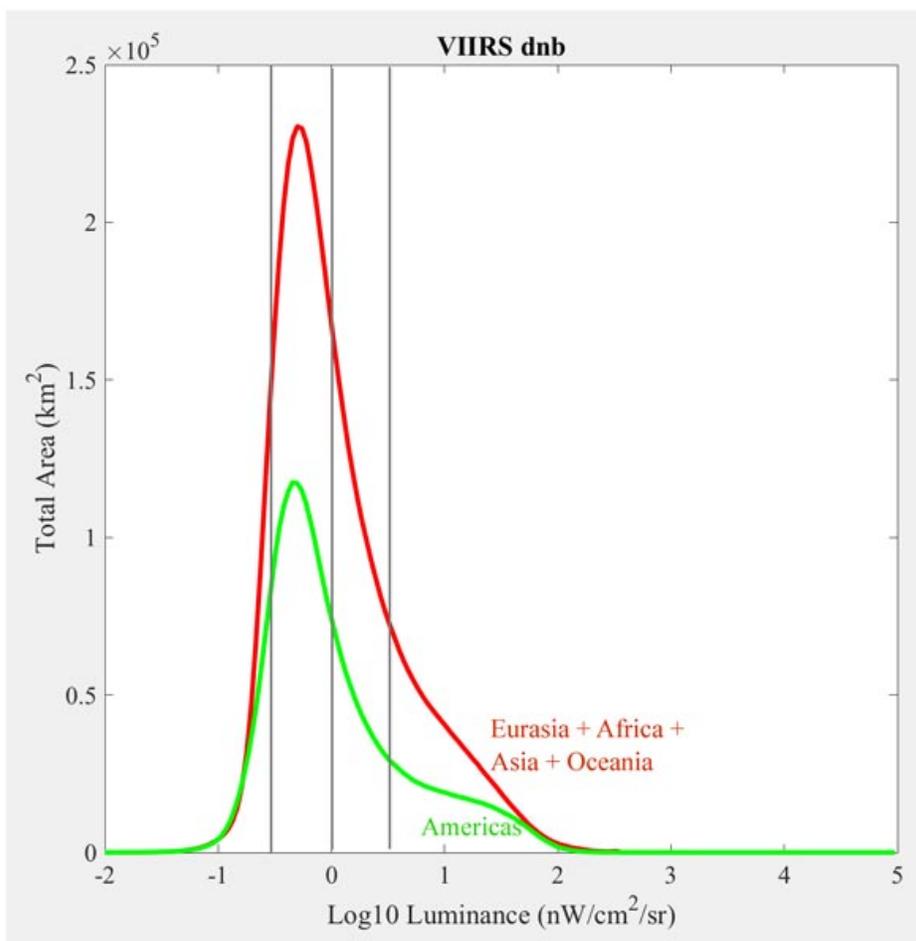

Figure 3 VIIRS dnb luminance distributions with thresholds superimposed. The lower (-0.5) cuts only the lower tail of dimmest pixels while the upper (+0.5) retains only the shoulder and upper tail of brightest pixels. The medium (0) retains approximately half.

The process of segmenting a continuous luminance field into discrete spatially contiguous subsets of pixels (segments) produces a spatial network in which each segment represents a distinct network component (subset of linked nodes). In each component, adjacent lighted pixels are analogous to nodes with links implied by adjacency.  Prior to segmentation, the VIIRS-masked DMSP composites are separated into three longitudinal subsets for 1) North, Central and South America, 2) Europe, Western Asia and Africa, 3) Central and Eastern Asia and Oceania.  The resulting masked composites are reprojected into Molleweide (1) and Sinusoidal (2 & 3) equal area projections at 500 x 500 m resolution.  Each masked DMSP composite is segmented with Queen's case adjacency neighborhood using an 8 DN threshold and a 25 pixel minimum segment size.  The 8 DN threshold is based on the OLS noise floor below which the frequency of spurious detections increases abruptly [*Small et al.*, 2011].

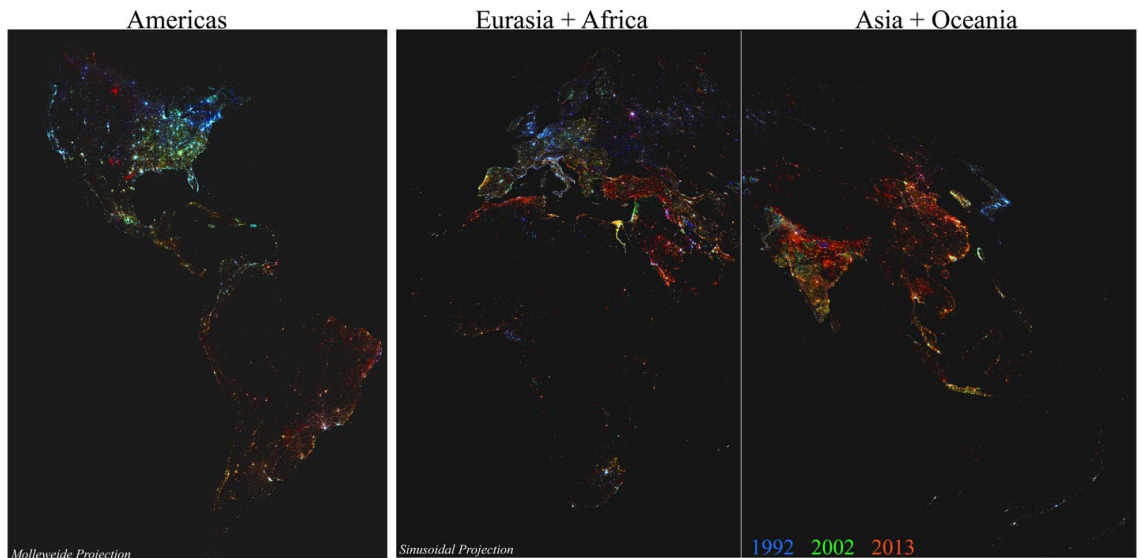

Figure 4  Tri-temporal DMSP+VIIRS composite.  Color implies change.  Warmer colors indicate brightening.  At global scale, stable urban cores are too small to be seen and change appears pervasive. In fact, most change is associated with large areas of infill development, rural electrification and expansive development.   The dividing line between Eurasia + Africa and Asia + Oceania is chosen to avoid breaking interconnected network components.

For each spatially contiguous segment, area and perimeter are calculated and sorted to produce rank-size and rank-shape distributions for a total of 27 combinations of 3 years x 3 regions x 3 thresholds.  The DMSP+VIIRS tri-temporal fused change product is shown in Figure 4 and the least conservative ($10^{-0.5}$ threshold) tri-temporal segment area composite is shown in Figure 5. Because segment areas (and perimeters) span 4+ orders of magnitude, the segment area composites show $Log_{10}$ area of 2013, 2002 and 1992 in the R,G,B channels respectively.  As with the tri-temporal brightness composite, color implies change and warmer colors (red, yellow) indicate growing network components. A linear stretch is applied to each year channel of the image so the components changing the least appear shades of gray, with larger components being lighter.

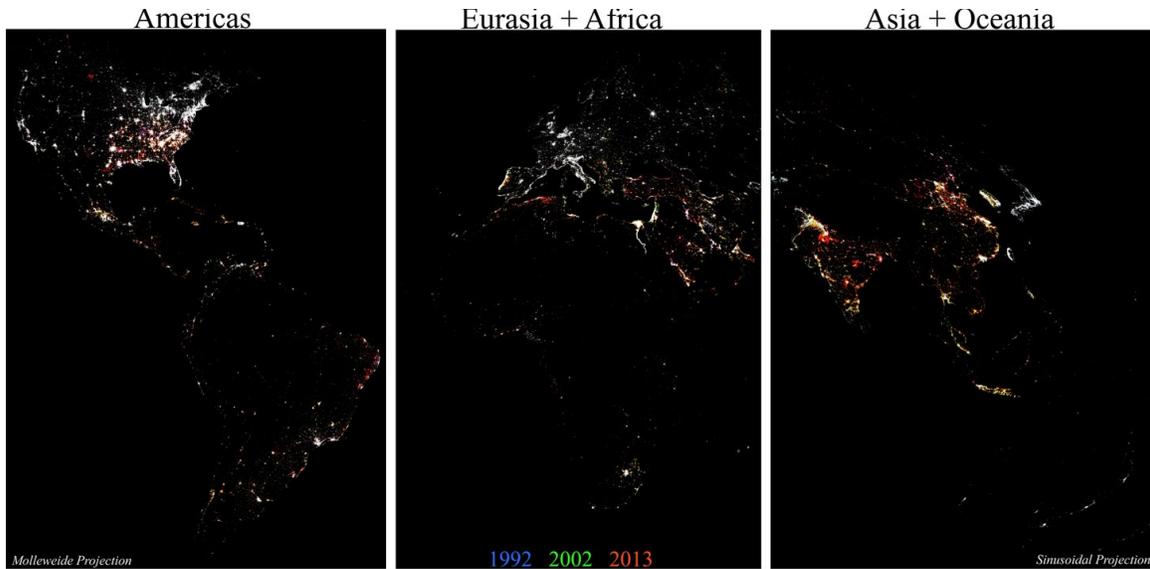

Figure 5 Tri-temporal DMSP+VIIRS network area composite. Brightness corresponds to $Log_{10}$ area of each spatially contiguous network component. Stable parts of network components appear white. Warmer colors indicate network growth. Because component areas span 4 orders of magnitude, only the largest are resolved at global scale. Full resolution examples in appendix show a wider range of component sizes.

## Results

Area-perimeter distributions of network components show the expected fractal scaling for all years and thresholds (Figure 6). Despite the quasi-linear scaling, there is a clear upward curvature in each distribution, indicating that the morphology of the network components becomes increasingly tortuous as they grow larger. Regression of $Log_{10}$ perimeters on areas give fractal dimension estimates in the range of 1.37 to 1.51 with correlations in the range of 0.94 to 0.97 for all 27 subsets. These dimensions are consistent with, if somewhat higher than the fractal dimensions estimated for intraurban land use categories by [*Batty and Longley*, 1994]. The implication is that the fractal structure observed for intraurban land use extends beyond the scale of individual settlements to describe the morphology of vastly larger spatial networks of development.

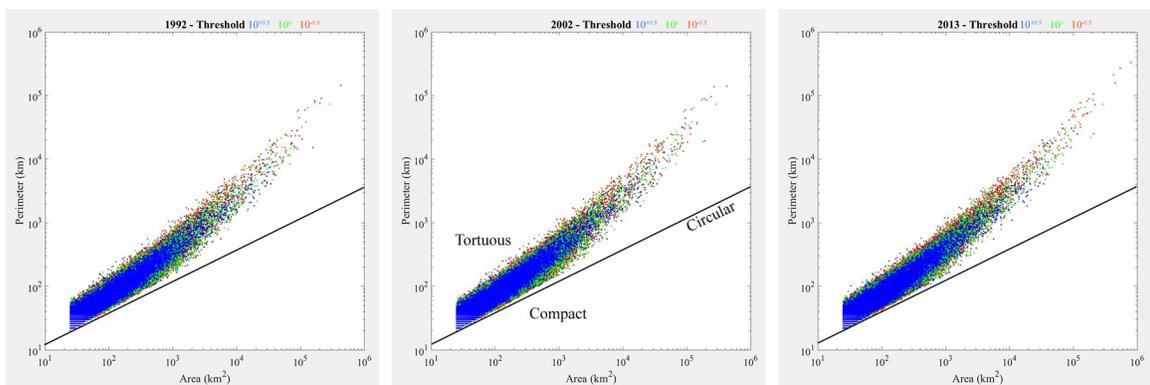

Figure 6 Global area-perimeter distributions. Despite the quasi-linear scaling over 4 orders of magnitude, there is a clear upward curvature and tapering of each distribution as the larger components become increasingly tortuous. Only the smallest components approach the circular scaling limit shown by the diagonal line.

Rank-size distributions are strongly linear and well-described by power laws with exponents within ±0.07 of −1 for all 27 subsets.  As shown in Figure 7, the high threshold ($10^{+0.5}$) distributions show a significant roll off for the largest components – despite having slopes slightly steeper (more negative) than −1.  Physically, this means that the higher threshold has the effect of fragmenting the largest components disproportionally.  This effect is most pronounced in North America, to a lesser degree in Eurasia, but not observed at all in Asia.  The high and intermediate thresholds produce very similar distributions for all three years in the Americas and Eurasia+Africa, reflecting the relative stability of the brighter urban cores.  In contrast, the intermediate threshold produces distinct and uniformly growing distributions for Asia.  The low threshold ($10^{-0.5}$) shows the network evolution most clearly in all regions, as entire distributions shift upward in parallel.  The effect of the low threshold is most pronounced for the largest components of the Asia subset.  The largest components of the Asian network show a clear upward departure from the linear trend between 2002 and 2013.

The rank-shape distributions show very similar progressions to the rank-size distributions (Figure 8).  As expected, the Asian network experiences the greatest change for all thresholds, particularly between 2002 and 2013.  The upward shift of the perimeter/√area distributions reflects the increasing tortuousity of network components of all shapes with time – as would be expected from the area-perimeter scaling seen in Figure 6.  As with the rank-size distributions, the most pronounced change is seen for the largest, most tortuous networks with the low threshold mask.

The different characteristics of the geographic subsets are less apparent when they are combined to form global rank-size and rank-shape distributions (Fig. 9).  The linearity and slopes of the rank-size distributions persists, as does the upper tail roll off of the low and intermediate threshold distributions.  The progressive upward shift of the largest, most tortuous, network components is more apparent for all thresholds of the global rank-shape distributions.  In the low threshold distributions, the 8 largest, most tortuous, components show a clear departure from the rest of the distribution in 1992 and 2002.  However, by 2013 three much more tortuous components have emerged, while the next 35 components also show a clear upward departure from the more linear continuum below.

Luminance distributions within network components show an abrupt transition with size of component.  Appendix Figures 10a, b, c show bivariate distributions of total network area as a function of component size and luminance for the lowest threshold (-0.5) networks.  For a continuum of smaller component sizes (x), luminance (y) is strongly skewed toward low values, with a long tail of brighter lights.  However, for components larger than 3000-5000 $km^2$ the distributions abruptly become discontinuous in size, with more uniform distributions of luminance for each of the larger components.  For the very largest components in America and Asia, the luminance distributions again become skewed toward low values indicating the abrupt inclusion of large areas of low luminance in the periphery of the largest components.  The geographic distribution of components in the discontinuous upper tails are shown in red superimposed on the background light distribution in Figure 10a, b, c. While the majority of smaller components correspond to individual cities of varying size, the largest represent interconnected networks of cities and their periurban surroundings.

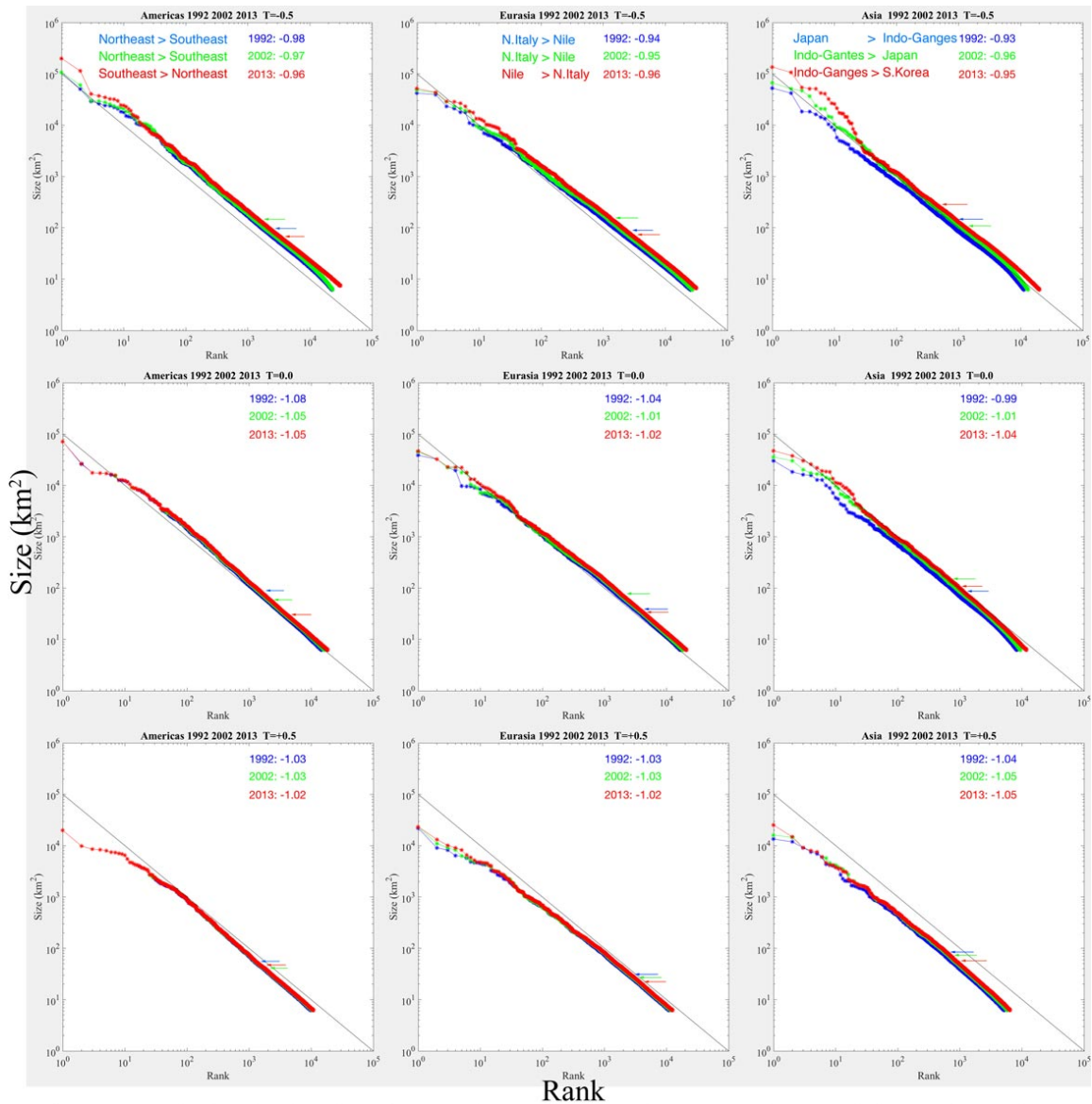

Figure 7 Rank-size distributions for 27 subsets of geography, year and connectivity. All have slopes within 0.08 of -1, but the uppermost tails show varying sensitivity to luminance threshold. Arrows show lower tail cutoff of minimum MLE misfit, although linearity clearly extends lower. Growth and interconnection result in displacements between two largest components in each quatrosphere.

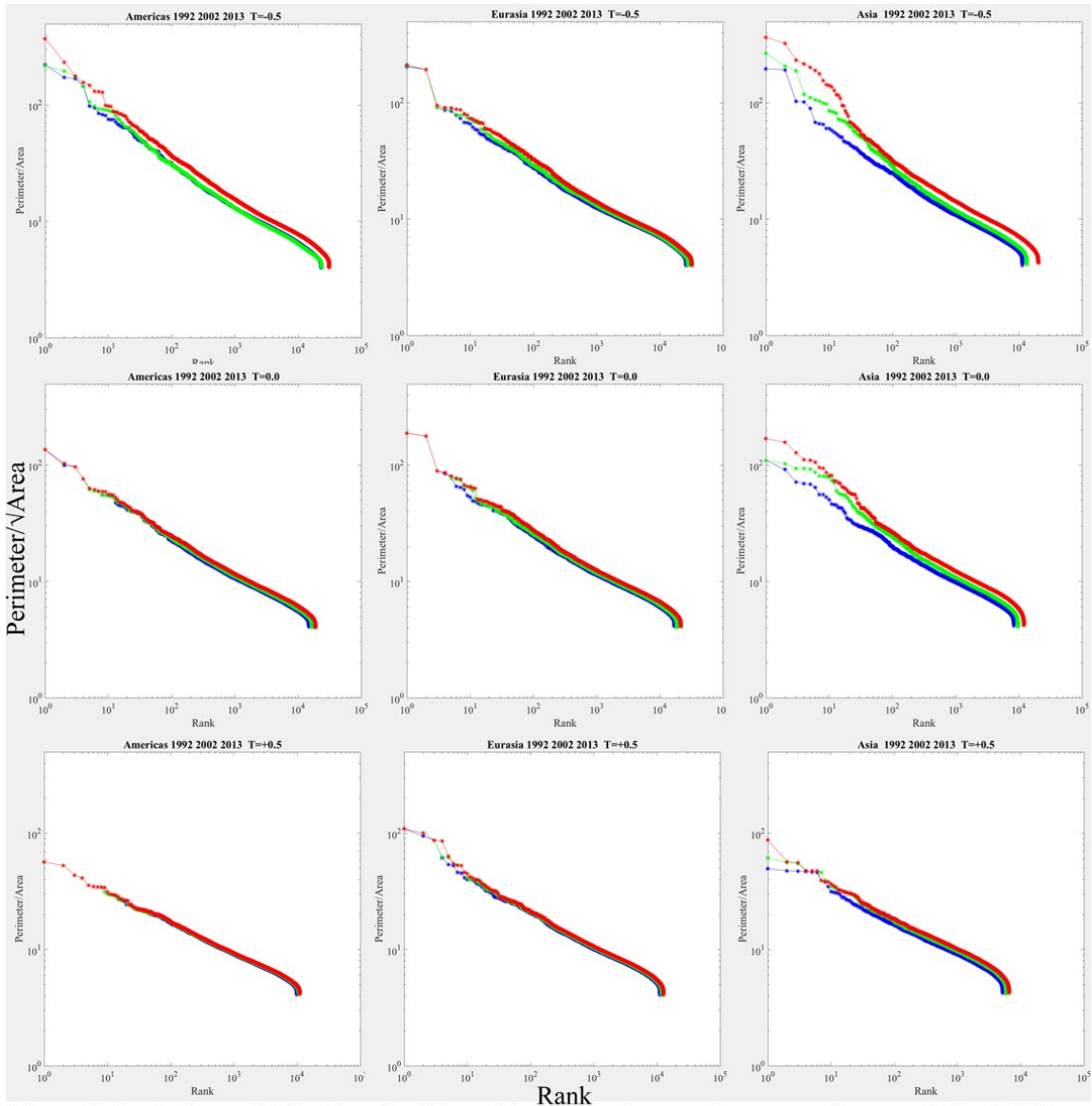

Figure 8 Rank-shape distributions for 27 combinations of geography, year and connectivity. While the stable cores (T=+0.5) show little change, at lower luminance threshold (T=-0.5) the largest, most tortuous network components show accelerating growth in Asia and the Americas.

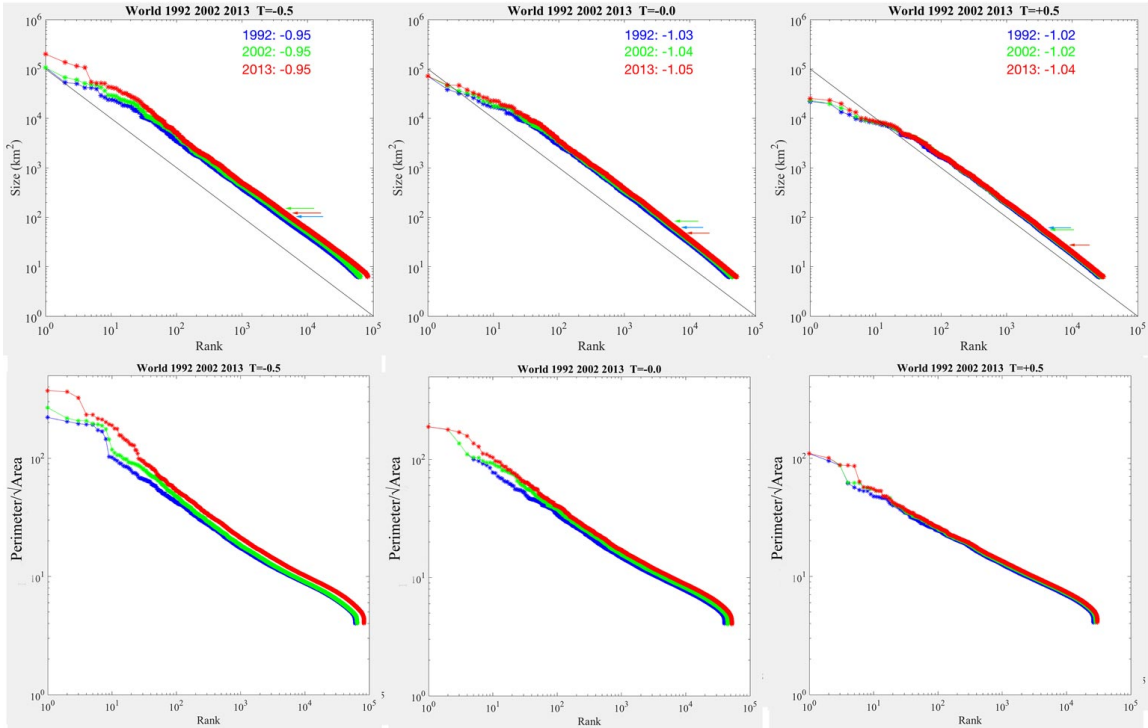

Figure 9  Global rank-size and rank-shape distributions. The uppermost tails of all the rank-size plots fall below the linear trends, but the most tortuous network components lie above the trend of the rank-shape plots, suggesting that diminishing increase in component size is offset by increasing morphologic complexity.

Network component size composites for some of the largest agglomerations illustrate the spatiotemporal evolution of these networks over the past two decades.  Appendix Figures 11a, b, c show tri-temporal luminance composites along with component size composites for the largest components in N. America, S. Asia and E. Asia respectively.  The color scheme is the same for both luminance and size with areas in white representing the stable component core present since 1992, yellow representing growth during the 1990s and red representing subsequent growth during the 2000s.  Larger components appear lighter with smaller components darkening logarithmically.  Figure 11a shows the Southeast Corridor of the USA, which surpassed the Northeast Corridor in size during the 2000s.  Interconnection of Atlanta, Charlotte and Raleigh-Durham along the eastern Piedmont of the Appalachian mountains in the 1990s, extended to Chattanooga and Knoxville on the western Piedmont during the 2000s.  The two largest components in south and east Asia are somewhat smaller than that in the US, but both grew much more rapidly in the 2000s, along with the next 10 largest in Asia departing upward from the linear trend of their rank-size distributions.  The largest component in Asia is on the Indo-Gangetic Plain extending from Delhi and the Punjab northward into the Indus River Valley in Pakistan in the 1990s and then southward into Rajasthan in the 2000s.  The largest component in China is comparable in size but very different in structure and evolution to the Indo-Gangetic Plain component.  The Huanghe Delta + North China Plain component was dominated by the Beijing-Tianjin corridor during the 1990s, with several smaller multi-city components to the south, but grew abruptly during the 2000s as parallel growth of the entire network resulted in simultaneous interconnection into a single, much larger component by 2013 (Fig. 11c). At the same time, the smaller Changjiang Delta + Yangtze River component to the south also grew rapidly during the 2000s and has almost connected to the North China Plain component. When these two components interconnect they will produce a single component much larger than that of the northern Indo-Gangetic Plain.

## Discussion

While none of the conclusions drawn here depend on the observed network scaling being explicitly power law, the consistency of the fits over 4-5 orders of magnitude in size and number is compelling. The presumption of power law scaling is also consistent with the generative conditions for power law network emergence from random networks generally. Whereas scale-free networks can emerge from random networks through the combined processes of growth and preferential attachment [*Barabási and Albert*, 1999], it has been shown that scale-free spatial networks can also emerge from random continuous fields through the same combined processes of growth and preferential attachment [*Small and Sousa*, 2015]. The process of progressive segmentation of a random continuous field naturally results in growth of a foreground network while the increase in perimeter of the growing components naturally results in preferential attachment to the larger components with the longer perimeters. In other words, the same conditions for emergence of scaling in random networks (growth & preferential attachment) generally can also explain emergence of scaling on bounded spatial networks. Growth is a consequence of progressive segmentation while preferential attachment is a consequence of growth within the confined domain of the lattice. The superlinear increase fractal dimension with component size demonstrated in Figure 6 effectively amplifies the preferential attachment mechanism.

The remarkably consistent scaling and evolution of the spatial networks described here has a number of implications and some potential applications.

### *Implications*

*The observation that network component morphologic complexity increases with size has implications for network growth processes.* For the smallest network components, the variation in morphologic complexity (tortuous vs compact) spans about one half order of magnitude in perimeter, but the upper bound of complexity spans more than three orders of magnitude from the smallest to the largest components (Fig. 6). This accelerating increase in the fractal dimension of larger component boundaries is the mechanism for preferential attachment that drives the consistent scaling behavior observed in all the networks considered here, and therefore gives rise to the observations and implications that follow.

*The observation that component scaling remains stable during network growth has implications for a quasi-equilibrium of lighted development worldwide.* Maintenance of rank-size distribution slopes near –1 over two decades of network growth worldwide implies a dynamic equilibrium between the underlying processes of nucleation, growth

and interconnection [*Small and Sousa*, 2015] that is maintained while network components grow at equal rates over most of their size distribution – as must occur to maintain consistent unity slope while growing.  Rank-size slopes vary by < 0.02 for seven of the nine networks considered here.  A unity-slope rank-size distribution can be understood as a transitional stage from increasingly balanced nucleation, growth and interconnection to rapid continuous percolation and formation of a Giant Component.

Treating the power law as a top-heavy probability distribution $p(x) = Cx^{-\alpha}$ allows for derivation of Lorenz curves giving the fraction of total network area, *A*, as a function of the component size, *S*, from largest to smallest as $A = S^{(\alpha-2)/(\alpha-1)}$ defined for values of α > 2  [*Newman*, 2006].  The value of 2 corresponds to the limit below which the mean of the probability distribution diverges. As the slope of the rank size distribution, s, is related to the power law exponent, α, by $s = -1/(\alpha-1)$, an α = 2 corresponds to a unity slope of –1 [*Li*, 2003].  In the context of spatial networks on finite area domains, α approaching the limiting value of 2 causes the exponent of the Lorenz function to approach 0 and therefore fraction A to approach 1.0, corresponding to the case of a single Giant Component encompassing the full spatial domain.  This is illustrated for evolution of random spatial networks generated from normal and lognormal ballistic deposition by [*Small and Sousa*, 2015] whereby as random networks grow, rank-size distributions become increasingly linear, with slopes approaching –1, immediately preceding continuous percolation and rapid formation of a Giant Component.  In the results presented here, the effect of increasing network connectivity (by lowering the luminance threshold) is similar to that observed from 20 years of growth for each geographic quatrosphere.  The entire rank-size distribution increases in area and number while maintaining linearity over most of its range, but the uppermost tail grows faster to approach the linear trend of the rest of the distribution (Fig. 7).  However, it is noteworthy that the highest connectivity (lowest threshold) networks for Asia emerged conspicuously above the linear trend by 2013, suggesting disproportionate interconnection growth of the largest ~30 components.

*The observation that the largest components are the most dynamic has implications for abrupt, yet potentially predictable growth scenarios.*  The largest components in the upper tails of both rank-size and rank-shape distributions change the most with time, space (geography) and connectivity.  Largely because of the outsized effect of interconnection among the larger components.  Increasing underscaling of the largest components with higher luminance thresholds replicates the underscaling widely observed in population-based city size distributions by numerous studies over the past 40 years, highlighting the opposing effects of administrative fragmentation and periurban network connectivity in disrupting or maintaining network structure.  However, the spatially explicit nature of the network characterization allows for the use of geospatial proximity analysis to identify imminent component interconnections under different future growth scenarios.  This has particularly interesting implications for Asia as the large networks in eastern China continue to interconnect.

*The observation that the Asian networks have surpassed Eurabian and North American networks in size and complexity, while the sub-Saharan African and South America*

*networks have yet to enter rapid growth and intereconnection phases (respectively) has implications for shorter and longer term growth scenarios.* Particularly in the context of resource shortages and surpluses that may inhibit and drive future growth. Western Europe, dominated by a few large network components, remained virtually unchanged at all scales from 1992 to 2015, while expansive growth occurred in southeastern Europe, Turkey, the Middle East and North Africa. In the Americas, infilling growth in the eastern US created and maintained the largest components with expansive growth in Mexico and Brazil driven by both urban development and rural electrification. With the exception of South Africa, sub-Saharan Africa remains largely unlighted, although ongoing rural-urban migration creates immense potential for future growth. By far, the greatest changes have occurred in southern and eastern Asia. Growth in South Asia has been expansive, with the emergence of several corridors forming large components, while growth in east and southeast Asia occurred through both expansion and infilling.

*The observation that growth and interconnection in all three quatrospheres together brings global rank-size distributions closer to unity slope has implications for future impact of urban expansion at global scales.* The progressive increase in network component fractal dimension with size suggests that much of the future growth of the largest, most interconnected components, may be increasingly space-filling rather than expansive. Globally, by 2015 the four largest components (>100,000 km$^2$) are approximately twice the size of next largest four. These four largest are in North America and Asia, and three of the four show at least as much infilling as expansive growth. Only the relatively stable Northeast Corridor of the US shows a conspicuously lower morphologic complexity (P/√A) than the other three of the largest four. Eurabia and North Africa show persistent uniform growth at all scales over the past decades, but their largest components are much smaller than those in Asia and North America.

### *Applications and Speculations*

*Disease transmission* - In post-lockdown conditions, with mobility severely curtailed, spatial diffusion of a pathogen within exposed network components assumes a greater role than in early stage transmission through air and surface transportation networks. However, post lockdown diffusion is conditioned on the seeding of network components before lockdown occurs. The continuous progression from small numbers of large components to large numbers of small components, combined with the tendency for transport hubs to occur in medium to large components implies the likelihood of large numbers of small components remaining unexposed after lockdown occurs. The implications of spatial network scaling for disease transmission are discussed in more detail by [*Small et al.*, 2020]

*Habitat loss and fragmentation* – The observation that settlements, agricultural land use and forests show similar spatial network scaling globally suggests complementary processes may explain simultaneous foreground and background network evolution [*Small and Sousa*, 2015]. The processes of nucleation, growth and interconnection have complementary analogues in attenuation, shrinkage and fragmentation. Emergence of random networks from progressive segmentation of normal and lognormal random fields shows complementary evolution of foreground and background networks [*Small and Sousa*, 2015]. If the scaling observed for spatial networks of forest patches is generally true of other non-anthropogenic landscapes, the complementarity of these network component distributions can provide hard constraints for simulation of habitat loss and fragmentation under different development and conservation scenarios.

*Future urban growth* – The observed progression of this spatial network scaling over the past 20 years, and the inference of its progression from the deeper past, may provide an explanation for the clustering expansion paradox in which humans have adapted to live in every terrestrial biome on Earth, yet the vast majority of the population remains strongly clustered onto a small fraction of the potentially habitable land area [*Small and Sousa*, 2016]. A corollary of this paradox is that the majority of the human habitable land area on Earth remains either uninhabited or sparsely inhabited by dispersed small settlements. The fact that many of these smaller, more isolated settlements are engaged in agriculture suggests environmental impact disproportionate to area, yet suggests that increasing rural electrification, and adoption of renewable non-grid energy will allow for identification of smaller settlements. The observation of widespread infill development, and expansion being driven by rural electrification more than expansion of development suggests that the spatial network scaling observed over the past decades may continue for some time into the future.

**Appendix**

Figures 10a, b, c show bivariate distributions of total land area as functions of network component size and luminance, illustrating the abrupt change from a continuous distribution to a discontinuous succession of larger components with more uniform luminance. Figures 11a, b, c show higher resolution examples of the tri-temporal composites of luminance and network component area for the three largest network components.

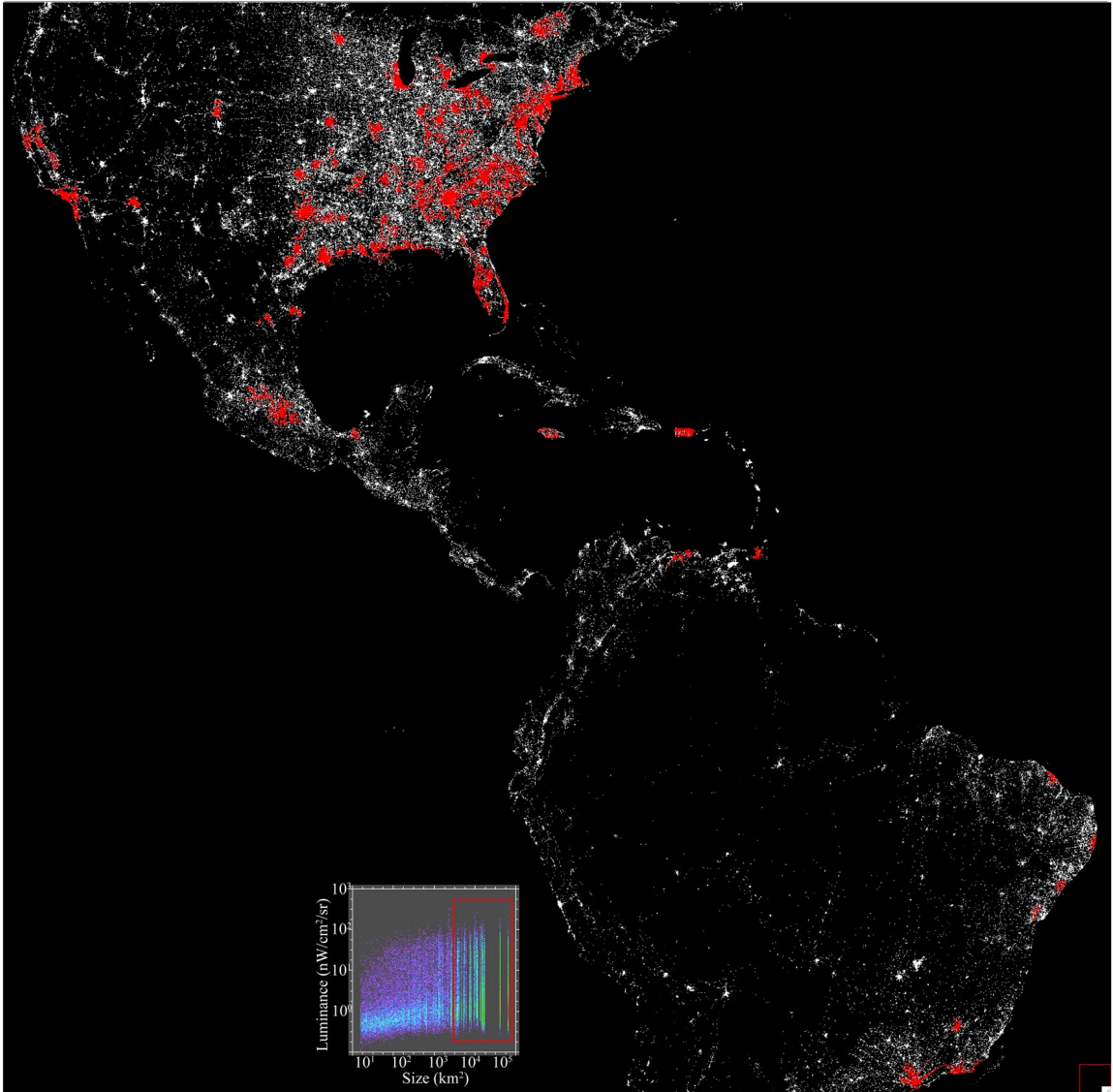

Figure 10a Size-luminance distribution and network component map for the Americas. The bivariate distribution of land area as a function of network component size and luminance (inset) shows an abrupt transition from a continuous, skewed distribution over many smaller components to a discontinuous series of larger components with more uniform luminance distributions. The discontinuous upper tail of the largest components (red box) of the low threshold ($10^{-0.5}$) network is shown in red against all VIIRS night lights (white). The network structure is dominated by the extensively developed eastern USA, although nascent subnetworks are apparent in central Mexico, and in the northeast and southeast of Brazil. *Enlarge to see fine detail.*

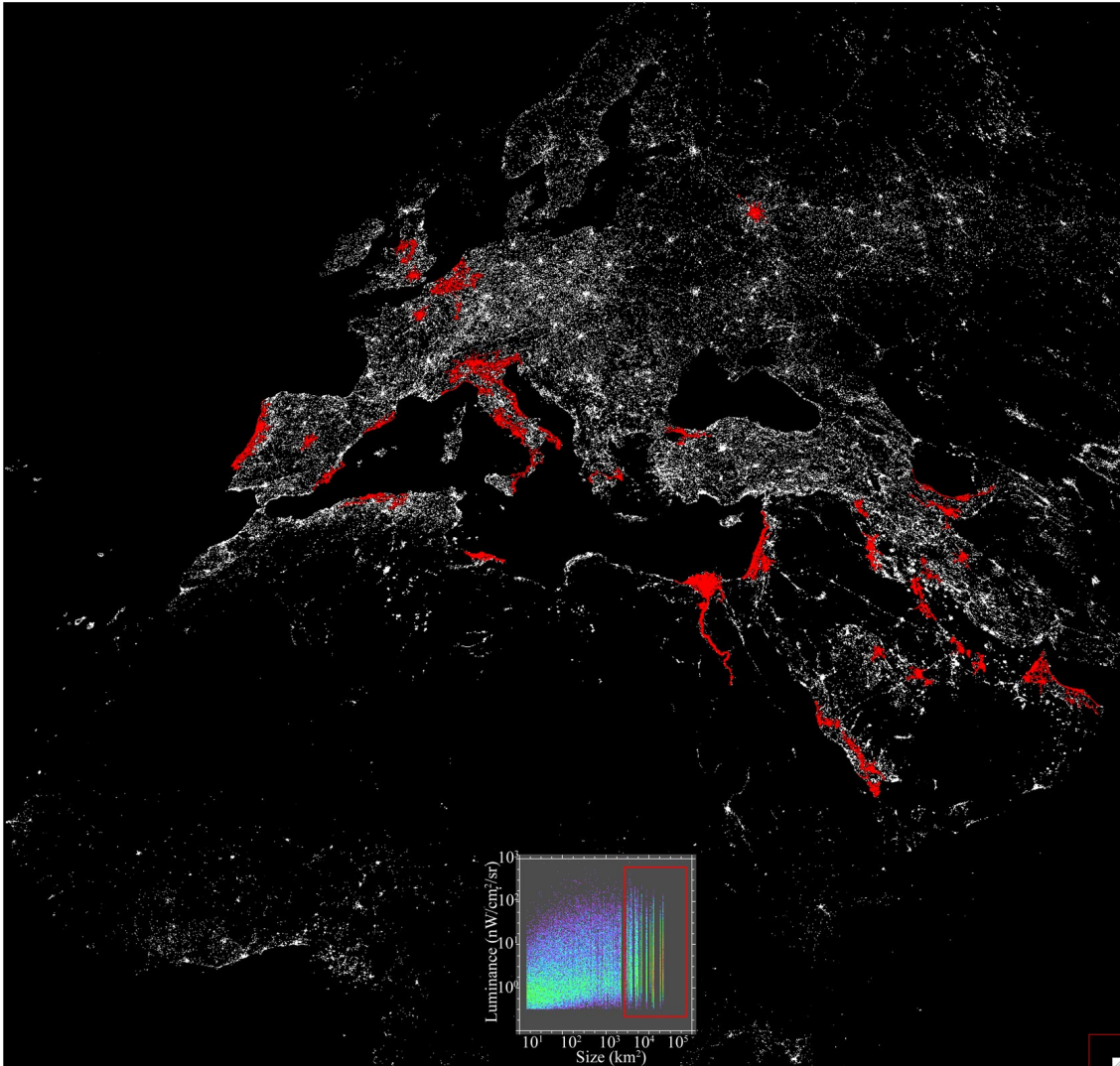

Figure 10b Size-luminance distribution and network component map for Eurasia and northern Africa.  Discontinuous upper tail of largest components (red box) of low threshold ($10^{-0.5}$) network shown in red against all VIIRS (white).  The largest component - the Nile Delta and river valley - is larger but more compact than the northern Italy component.

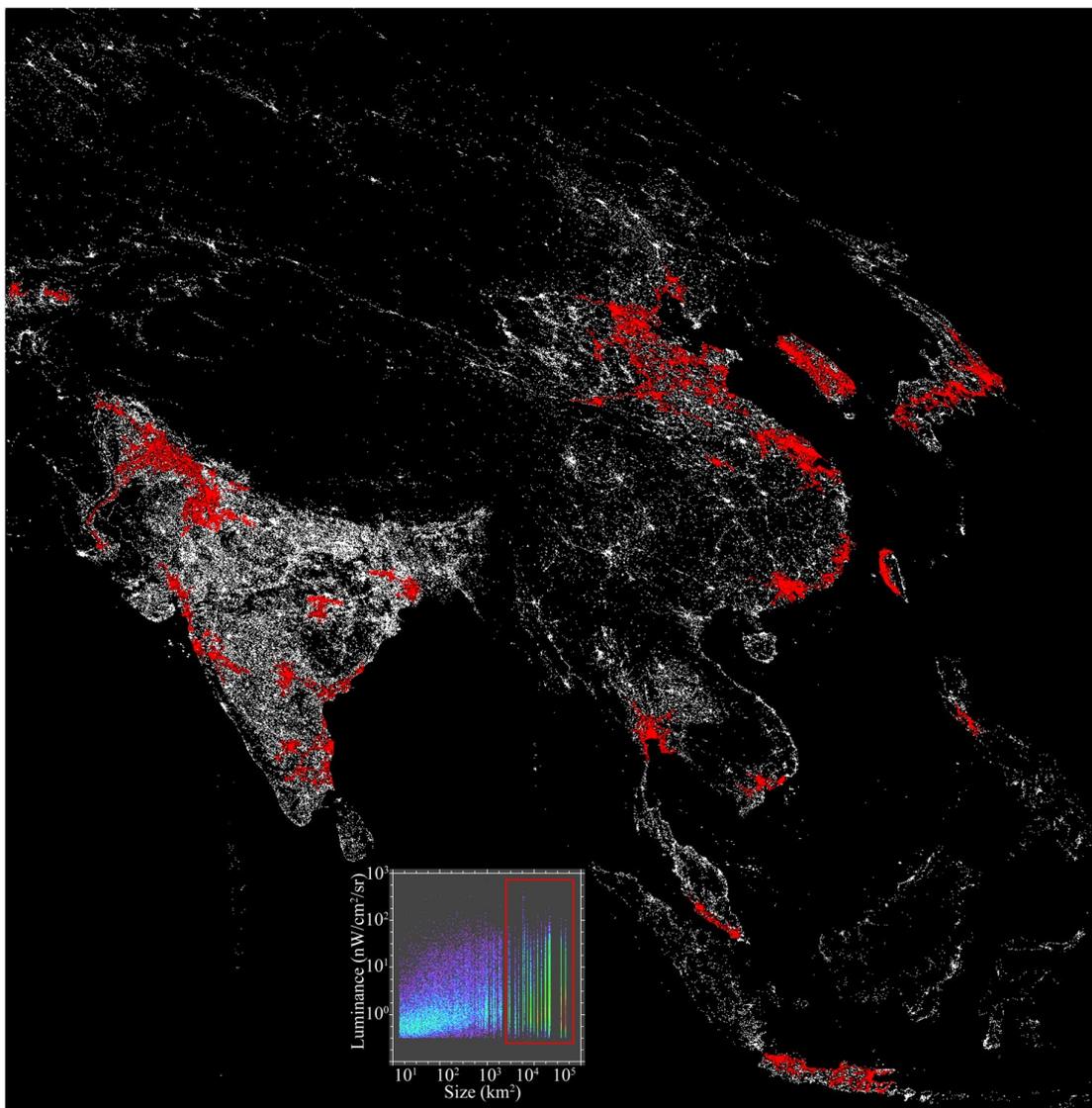

Figure 10c Size-luminance distribution and network component map for Asia. Discontinuous upper tail of largest components (red box) of low threshold ($10^{-0.5}$) network shown in red against all VIIRS (white). The Indo-Ganges and S. Korean components have recently displaced the Japan component to become the two largest in Asia. However the large components in east China are nearly connected.

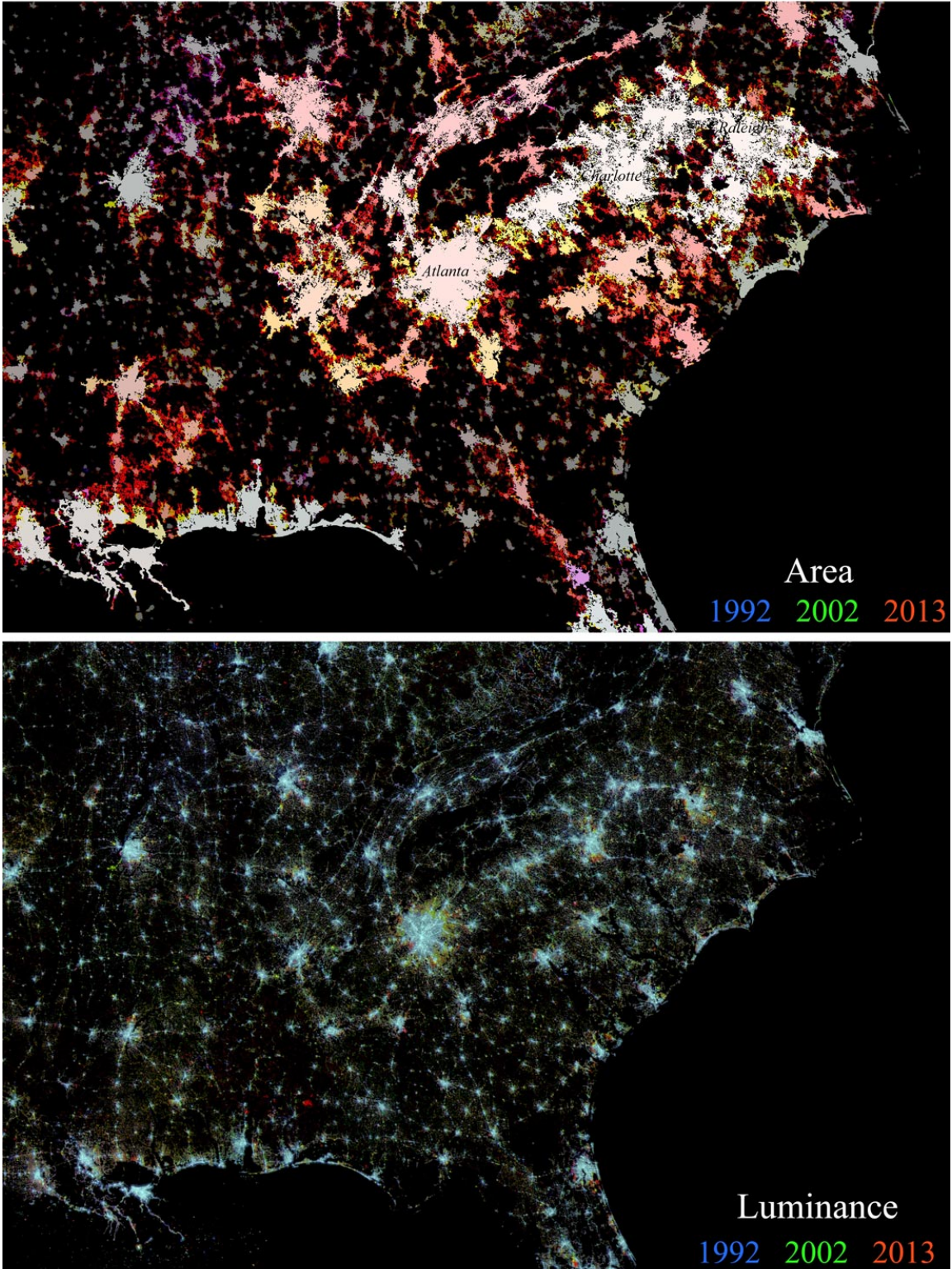

Figure 11a Network size composite and DMSP+VIIRS tri-temporal composite for the Southeastern USA. Since 2002, the Southeast Corridor (Atlanta-Charlotte-Raleigh-Durham) has expanded both east and west to surpass the more compact Northeast Corridor (Washington-Philadelphia-New York-Boston).in area

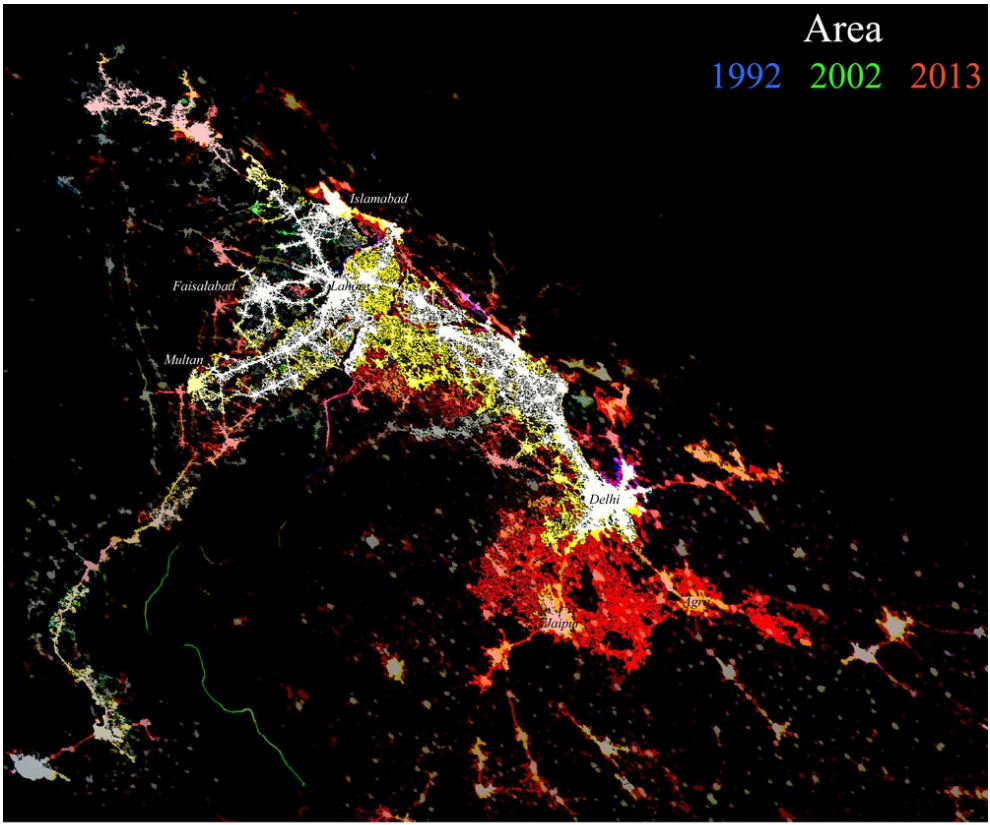

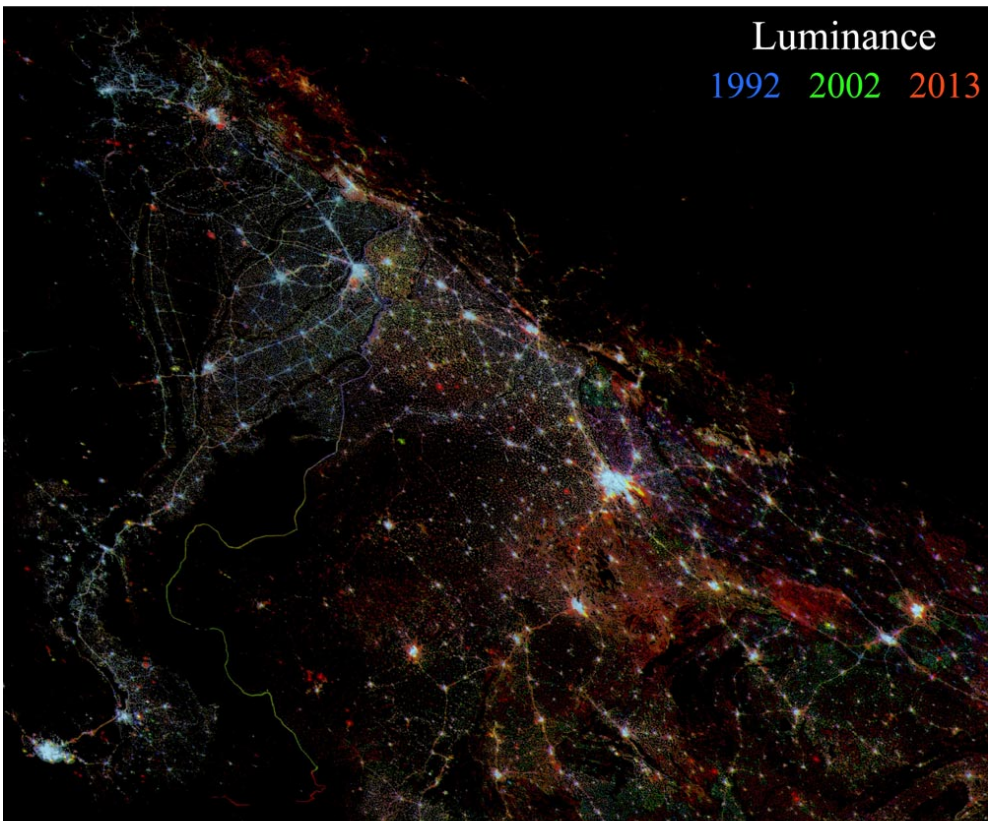

Figure 11b  Network size composite and DMSP+VIIRS tri-temporal composite for the northern Indo-Gangetic plain.  Post-1992, most growth was along the Lahore-Delhi axis, but post-2002 expansion has occurred faster to the south and east of Delhi.

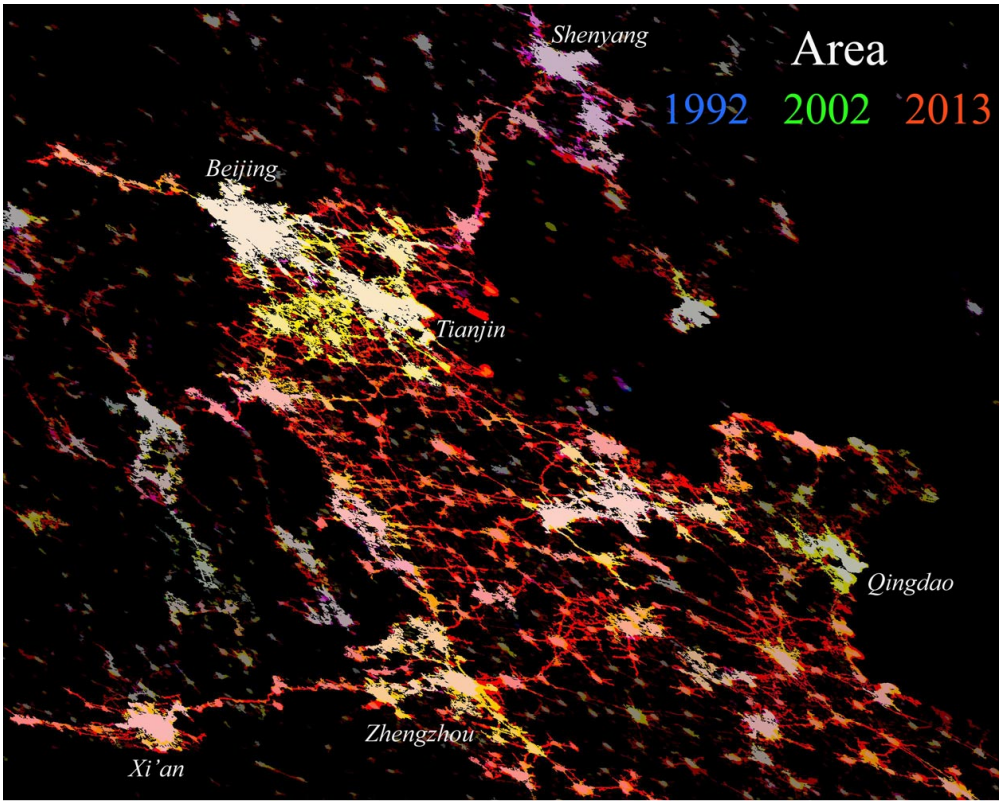

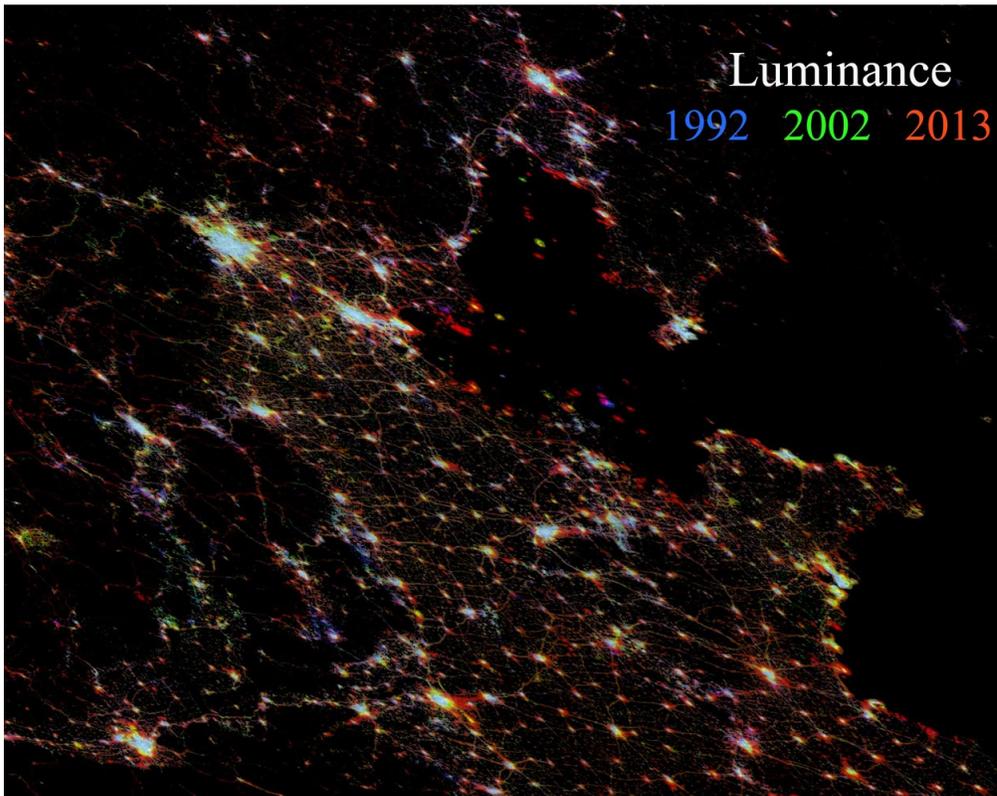

Figure 11c Network size composite and DMSP+VIIRS tri-temporal composite of the HuangHe delta + North China Plain. Post-2002, extensive interconnection of the cities of the plain has resulted in > 10 fold increase in area of the largest network component in China. Imminent connection to the second largest, the Changjiang Delta-Yangtze River Valley conponent to the south, will surpass the North Gangetic Plain to produce the largest component in Asia.

**Acknowledgements**

CS gratefully acknowledges funding provided by the Fulbright Foreign Scholarship Program and the Brazilian Federal Agency for Support and Evaluation of Graduate Education.